\documentclass[12pt]{article}

\usepackage{vmargin}

\setpapersize{A4}
\setmarginsrb{2.5cm}{2cm}{2.5cm}{2.5cm}{1.5cm}{0cm}{.5cm}{1.4cm}

\newcommand{\comment}[1]{}

\usepackage{pifont,amssymb}
\newcommand{\es}{\mbox{\Pisymbol{psy}{198}}}
\newcommand{\GSt}{{\sc STP[1,2] }}
\newcommand{\St}{{\sc STP[1,2] }}
\newtheorem{lemma}{Lemma}
\newtheorem{theorem}{Theorem}
\title{\bf 1.25 Approximation Algorithm for\\ the Steiner Tree Problem
with Distances\\ One and Two\vspace{0.4cm}}
\author{Piotr Berman\thanks{Department of Computer Science \& Engineering,
Pennsylvania State University, University Park, PA 16802. Research partially
done while visiting Department of Computer Science, University of Bonn and
supported by DFG grant Bo 56/174-1. Email: berman@cse.psu.edu}
\and Marek Karpinski\thanks{Department of Computer Science, University of
Bonn, 53117 Bonn. Supported in part by DFG grants, Procope grant 31022,
and the Hausdorff Center grant EXC59-1. Email: marek@cs.uni-bonn.de}
\and Alex Zelikovsky\thanks{Department of Computer Science, Georgia State
University, Atlanta, GA 30303. Email: alexz@cs.gsu.edu}
}
\date{}

\begin{document}
\maketitle
\abstract{
We give a 1.25 approximation algorithm for the Steiner Tree Problem
with distances one and two, improving on the best known bound for that problem.\\[1ex]
}
\section{Introduction}
We give a new approximation algorithm for the problem of finding a minimum
Steiner tree for metric spaces with distances one and two.  It improves
over the best known approximation factor for that problem of 1.279 \cite{RZ}.  Moreover,
unlike the result of Robins and Zelikovsky, our methods yields a single
algorithm, whereas \cite{RZ} gives an approximation scheme.

\section{Definitions and Notation}

A metric with distances 1 and 2 can be represented as a graph, so
edges are pairs in distance 1 and non-edges are pairs in distance 2.
We will denote by \GSt the Steiner Tree Problem restricted
to such metrics.

The problem instance of \St
is a graph $G=(V,E)$ that defines a metric in this way, and a set $R\subset V$
of {\em terminal nodes}.

A valid solution is a set unordered node pairs $T$ such that $R$ is
contained in a connected component of $(V,E)$.  We minimize $|T\cap E|+2|T-E|$.

An $s$-star of $G$ consists of a non-terminal $c$, called the center, $s$ terminals
$t_1,\ldots,t_s$ and edges $(c,t_1),\ldots,(c,t_s)$.
If $s<3$ we say that the star is {\em degenerate}, and {\em proper} otherwise.

In the analysis of the algorithm, we view the algorithm selections as transformations of the
input instance, so after each phase we have a partial solution and a residual
instance.  We formalize these notions as follows.

A partition $\Pi$ of $V$ induces a graph $(\Pi,E(\Pi))$ where
$(A,B)\in E(\Pi)$ if $(u,v)\in E$ for some $u\in A,v\in B)$.
We say that $(u,v)$ is a representative of $(A,B)$.

Similarly, $\Pi$ induces the set of terminals
$R_\Pi=\{A\in\Pi:A\cap R\not=\es\}$.

In our algorithms, we augment initially empty solution $F$.
Edge set $F$ defines partition $\Pi(F)$ into connected components of $(V,F)$.
In a step, we identify a connected set $X$ in the induced graph
$(\Pi(F),E(\Pi(F)))$ and we augment $F$ with representatives of edges
that form a spanning tree of $X$.  We will call it ``collapsing $X$'',
because $X$ will become a single node of $(\Pi(F),E(\Pi(F)))$.

\section{Analyzing Greedy Heuristics}
\label{ARSh}

We introduce a new way of analyzing greedy heuristics for our problem,
and in this section we illustrate it on the example of Rayward-Smith heuristic \cite{R-S}.
This heuristic has approximation ratio of exactly $4/3$, as demonstrated
by Bern and Plassman \cite{BP}.  However, the new analysis method will
allow us to see the effect of more general classes of greedy choices,
as we will show in the next section.

We will analyze now the following three steps algorithm.\\[1ex]

\fbox{
\parbox[c]{5in}{
\begin{description}
\item[{\bf Preprocessing:}]
If there is an edge $e\subset R$, we collapse $e$.  This increases the
hidden cost of 1, but it also decreases the cost of any solution by 1.

\item[{\bf Greedy selection loop:}]
Find an $s$-star $S$ with the largest possible $s$.  If $s=2$, exit the loop.
Collapse $S$.

\item[{\bf Finishing:}]
Connect the remaining terminals with non-edges (pairs with cost 2).
\end{description}
}
}\\[1ex]

In the analysis we use the following notions:

$~~~CA$, the cost of the partial solution formed by the algorithm;

$~~~CR$, the cost of the reference solution (derived from
an optimum);

$~~~P$, the sum of potentials distributed among objects.

\noindent
We initialize $CR$ to be the cost of an optimum solution, $CA$ is initialized
to $0$ (the algorithm starts with the empty partial solution) while the initial
potential that satisfies $P<CR/4$.  Later we need to show that

(a) at every step  $CA+CR+P$ is unchanged or decreased.

(b) the final value of $CR+P$ is zero.

\noindent
To show (a), the analysis of a step contains calculation of the {\em balance},
$-(\Delta CA+\Delta CR+\Delta P)$, and we need to show that balance is
nonnegative.

The initial reference solution is an optimum solution $T^\ast$.
The steps of normalization (performed only in the analysis) and of the
algorithm will alter the graph and $T^\ast$ until $T^\ast\cap E=\es$.
At that point, the Finishing finds optimal set of connections, as
all of them have cost 2.

The potential is given to the following objects:
\begin{itemize}
\item edges, elements of $T=T^\ast\cup T$
\item
{\em C-comps} which are connected components of $(V,T)$;
\item
{\em S-comps} which are Steiner components, or subtrees of  $T$
(connected subgraphs) in which leaves (nodes of degree 1) are terminals,
and internal nodes (of degree larger than 1) degree are non-terminals.
\end{itemize}
Initial potential assignment is $1/4$ to each edge and zero to other objects.

We will use $PC$ ($PE$, $PS$) for the sum of potentials of C-comps (edges,
S-comps).

We normalize the $T$ using steps with non-negative
balance which preserve the following additional invariants:
\begin{dingautolist}{172}
\item
each edge $e\in T$ has $p(e)=1/3$, and each C-comp $C$ with less than
two edges has $p(C)=0$.
\item
each C-comp $C$ has $p(C)\ge -2/3$, and each S-comp $S$ has $p(S)=0$.
\end{dingautolist}

\noindent{\bf Path Step:}
$T$ contains a path with $k>1$ edges between terminals or nodes of
degree larger than 2.  We remove this path from $T$
(implicitly, it adds a non-edge
to $T^\ast$).
As a result, $\Delta CR=2-k$ and $\Delta PE=-k/3$,
while we split a connected component $C$ into $C_0$ and
$C_1$; we set $p(C_0)=p(C_1)=0$, which yields $\Delta PE\le 2/3$.

The balance of Path Step is at least $k-2+k/3-2/3=2/3(k-2)\ge 0$.

\noindent{\bf Bridge Step:}
the Path Step does not hold and there is an edge between non-terminals.
We remove this edge from $T$ (again, this adds a non-edge to $T^\ast$).
$\Delta CR=1$ and $\Delta PE=-1/3$, while we split a connected component $C$
into $C_0$ and $C_1$.  Because the Path case does not hold and we have
no non-terminal leaves, $e(C_0)\ge 2$ and $e(C_1)\ge 2$.
We set $p(C_0)=p(C)$ and $p(C_1)=-2/3$, so $\Delta PC=-2/3$.

The balance of the Bridge Step is $-1+1/3+2/3\ge 0$.

The proof of the following is left to the reader:
\begin{lemma}
\label{afterCase}
After applying Path and Bridge steps, invariants \ding{172} and \ding{173}
hold true, and every Steiner component is either an $s$-star or
a single edge between two terminal nodes.
\end{lemma}

Now we will prove

\begin{lemma}
\label{afterPrep}
After Preprocessing, and after each step of Greedy selection with an $s$-star
where $s>3$, invariants \ding{172} and \ding{173} hold true and every S-comp
is an $s$-star.
\end{lemma}

\noindent{\bf Proof.}
The only change to S-comps during the Preprocessing is that
we collapse S-comp that consist of one edge only, so we need to show
that if we insist on invariants \ding{172} and \ding{173} a Preprocessing step
has a non-negative balance.  Such a collapses an edge between two terminals.
As a result, $\Delta CA=1$, $\Delta CR=-1$, $\Delta PE=-1/3$ (because the
potential of this Steiner component, which was $1/3$, is removed)
while $\Delta PC=0$.  Only $\Delta PC = 0$ requires some argument,
because we may create a C-comp with $e(C)<2$, which forces $p(C)=0$.
However, this would mean that $C$ consisted solely of edges between terminals
and such a component cannot be created by the Bride Case, so it had $p(C)=0$
from the time it was created.
Thus the balance is $-1+1+1/3>0$.  

Now consider a greedy selection of an $s$-star $A$ with $s>3$.
Its terminals are in some $a$ C-comps.  To break cycles created in $T^\ast$
when we insert the star, we remove $s-1$ connections, of which
$a-1$ are non-edges (that connect different C-comps), so we remove
$a-1$ non-edges and $s-a$ edges.

Based on this characterization,
$\Delta CA=s$, $\Delta CR=-s-a+2$, $\Delta PE=(s-a)/3$.  Because we may reduce
the number of C-comps by $a-1$, we may have $\Delta PC=(a-1)2/3$,
and this leads to the balance of $(s-4)/3\ge 0$.

The balance calculation is false if we create a single C-comp with fewer than 2
edges, and this means, a single node.   In this case the resulting C-comp must
have potential $0$, so where we had perhaps $a$ C-comps with potential $-2/3$
each now we have one with potential 0 and $\Delta PC=a2/3$.   This exceed our
initial estimate by $2/3$.  But in this case we also alter the estimates
of $\Delta PE$ and $\Delta CR$.  In particular,
in at least one $s$-star we removed a maximum
number of edges, $s-1$, that is justified by the number of cycles we need
to break; however, the last edge is unnecessary (it creates a non-terminal
leaf in the reference solution), so we can simply remove it. 
This removal decreases $CR$ by $1$, and $PE$ by
$1/3$, so the balance estimate can be changed by $-2/3+1+1/3=2/3$,
in other words, in this case the balance is better rather than worse.

To finish the proof, we need to eliminate from $T$ S-comps that
are mot proper stars.
Initially, degenerate stars can be created by edge
deletions in proper stars.  As described above, a 1-star can be simply removed
which improves the balance.  A 2-star can be eliminated using a Path Step.

When we perform selections of 3-stars, \ding{173} is no longer preserved,
instead we will have invariant
\begin{dinglist}{203}
\item
each C-comp $C$ has $p(C)\ge -1/2$, and each S-comp $S$ has $p(S)\ge -1/6$.
$p(S)\ge -1/6$.
\end{dinglist}

The behavior of the algorithm during the selection of 3-stars can be
characterized as follows:
\begin{lemma}
\label{threestar}
After the last Greedy Selection with an $s$-star such that $s>3$ and
after each step of Greedy Selection with a 3-star invariants \ding{172}
and \ding{203} hold true and every S-comp is an $3$-star.
\end{lemma}

After the last selection of a star larger than 3-star we clearly have no
more S-comps that are different from 3-stars, because they would be
larger stars and would be selected.  Also, invariant \ding{173} implies
\ding{203} because we can increase the potential of each
non-trivial $C$-comp by $1/6$ and decrease the potential of one of
its S-comps from $0$ to $-1/6$.

Thus it remains to show that a selection of a 3-star
has a non-negative balance when we enforce \ding{172} and \ding{203}.

Because we select a 3-star, $\Delta CA=3$.

Suppose that the terminals of the selected star belong to 3 different
C-comps.  Then we remove 2 non-edges from $T^\ast$, so
$\Delta CR=-4$, we do not change S-comps so $\Delta PE+\Delta PS=0$,
while we coalesce 3 C-comps into one, hence $\Delta PC=1$, and the balance
is 0.

Suppose that the terminals of the selected star belong to 2 different
C-comps.  $\Delta CR=-3$ because we remove one non-edge
connection from $T^\ast$ and one edge from an S-comp.  This S-comp
becomes a 2-star, hence we remove it from $T$ using a Path Step, so
together we remove 3 edges from $T$ and $\Delta PE=-1$ and $\Delta PS=1/6$.
Because we coalesce
two C-comps, $\Delta PC=1/2$.
Thus the balance of the step is $1/3$.

If the terminals of the selected star belong to a single C-comp and we remove
2 edges from a single S-comp, we also remove the third edge of this S-comp and
$\Delta CR=-3$, while $\Delta PE=-1$, $\Delta PS=1/6$, and if its C-comp
degenerates to a single node, we have $\Delta PC=1/2$ (otherwise, zero).
Thus the balance is at least $1/3$.

Finally, if the terminals of the selected star belong to a single C-comp and
we remove 2 edges from two S-comps, we have $\Delta CR=-2$.  Because we apply
Path Steps to those two S-comps, $\Delta PE=-2$.  while $\Delta PS=1/3$ and
$\Delta PC\le 1/2$.  Thus the balance is at least $1/6$.

We formulate now

\begin{theorem}
The approximation ratio of Rayward-Smith heuristic is 4/3.
\end{theorem}

We initialized $CA=0$, $CR=opt$ and $P\le opt/3$, the sum $CA+CR+P$
never increases, and when we finish star selection we have
$P=0$ and $T=\es$.  At this point, we can connect the partial solution
using exactly the same number of non-edges as we have in $T^\ast$ so
we obtain a solution with cost $CA+CR$.

\section{A 5/4 Approximation Algorithm for \St}

We will formulate a new approximation algorithm
and we analyze it in the manner introduced in the previous section.

We generalize a notion of a {\em star}.  A {\em comet} is an S-comp
(a Steiner component) in which a non-terminal node is the {\em center},
center is connected terminals as well as to $a$ non-terminals called
{\em fork nodes}; each fork node is connected only to the center
and two terminals, those three nodes and edges form a {\em fork}.
Pictorially, a comet is like a star with trailing tail consisting of forks.
A comet in which the center is connected to $b$ terminals is
called a $b$-comet, or $(a,b)$ comet if it has $a$ forks.

The new {\em Six-Phase Algorithm} proceeds as follows:\\[1ex]

\fbox{
\parbox[c]{3.5in}{
\begin{enumerate}
\item Greedily collapse edges between terminals.
\item Greedily collapse $s$-stars with $s>4$.
\item Greedily collapse $s$-stars with $s=4$.
\item Find a maximum size set of 3-stars.
\item Greedily, replace a 3-star with a (1,3)-comet.
\item Greedily, collapse a comet with the least cost index.
\end{enumerate}
}
}\\[1ex]

We define the cost index $ci(S)$ of an
S-comp with $t$ terminals and $c$ edges as
$ci(S)=c/(t-1)-1$.  One can easily see that

\begin{lemma}
$$\mbox{If $S$ is an $s$-star, } ci(S)=\frac{1}{s-1},
\mbox{ if $S$ is an $(a,s)$-comet }
ci(S)=\frac{a+1}{2a+b-1}.
$$

\end{lemma}

It is somewhat less obvious that
\begin{lemma}
We can find a star or comet $S$ with the minimum $ci(S)$ in polynomial time.
\end{lemma}
First, observe that if $b\ge 3$, the cost index of a $b$-comet
does not increase if we delete the forks 
Thus we first check for a largest star.

When $b\le 2$ the situation is opposite: $(a+1,b)$-comet has lower
cost index than an $(a,b)$-comet.  Thus we can find the best
comet as follows.  Try every possible center, and for
each center create a graph in which edges are pairs of terminals
that can be connected to a single fork node which in turn is connected
to the center (disregard terminals directly connected to the center).
In this graph find a maximum matching.

In the analysis, we use similar potential as in Section \ref{ARSh},
with objects that get potential defined by the reference solution $T^\ast$
and $T=T^\ast\cap E$.  We define S-comps and C-comps as before, and
we initialize the analysis as before, with $T^\ast$ being the optimum
solution, except that we have $p((e)=1/4$ for $e\in T$.

\subsection{Analysis of Phases 1 and 2}

As in Section \ref{ARSh}, collapsing edges between terminals have a positive
balance.

Collapsing an $s$-star for $s>4$ removes $s-1$ edges from $T$,
so $\Delta C\le 1$ and $\Delta P\le -(s-1)/4\le -1$.

\subsection{Preliminary to the Analysis of Phases 3-6}

We normalize the reference solution using Bridge Steps and Path Steps.
The difference from Section \ref{ARSh} is that when we consider edge $e$ in
a Bridge Step, and removing it from $T$ would split some C-comp $C$
into $C_0$ and $C_1$, we (a) require that $|C_i|\ge 3$, $i=0.1$ and
(b) while one of $C_i$'s inherits $p(C)$, the other gets
$p(C_i)=-3/4$.

\begin{lemma}
After normalizing the reference solution $T^\ast$ using Bridge and Path steps,
every nontrivial S-comp is either an $s$-star, $s>2$ or
an $(a,b)$-comet, $a+b>2$.  Each S-comp will have potential 0, and 
each C-comp, zero or $-3/4$.
\end{lemma}

The proof is obvious.

\subsection{Analysis of the Phase 3}

Now we discuss the phase of selecting 4-stars.  We change the potential
distribution by setting $p(C)=-2/3$ for each C-comp $C$ that had potential
$-3/4$, and we compensate by setting, for one of its S-comps, say, $S$,
$p(S)=-1/12$.

When we select a 4-star, we remove 3 connections from $T^\ast$.  
To each such removal we can ``allocate'' $4/3$ of $\Delta CA$ (i.e. we
split the cost of the selected 4-star into three equal parts).

When we remove a non-edge, we have $\Delta CR=-2$, so $\Delta C=-2/3$.
We also coalesce two C-comps, so we may to increase $PC$ by $2/3$,
(cancel one of $p(C)=-2/3$).  Edges and S-comps
are not affected, so we get a balance.

When we remove an edge from a fork (i.e. incident to a fork node of
a comet), we can apply Path Step and remove two more
edges from $T$.  Thus we have $\Delta C=1/3$ and $\Delta PE=-3/4$,
a positive balance.  The balance is even better when we remove two
edges from the same fork, because in that case $\Delta C$ is better;
in the solution we erase three edges of a fork, rather then erasing
one and replacing two with a non-edge.  Thus we have 
$\Delta C=8/3-3=-1/3$, and we have $\Delta PE=-3/4$.

When we remove an edge from a 3-star or a ``borderline'' comet,
like $(2,1)$-comet or $(3,0)$-comet, the reasoning is similar to
the fork case.  We have significant surplus.  We also eliminate a
negative potential of the star, but the surplus is so big we will not
calculate it here.

The cases that remain is removing edges from stars, or edges
that connect terminals with centers of comets.  Without
changing the potential of the affected S-comp we would have a deficit:
$\Delta C=1/3$ and $\Delta PE=-1/4$, for the ``preliminary'' balance
of $1/12$.  We obtain the balance by decreasing the potential
of the affected S-comp by $1/12$.

This process has the following invariants:

\begin{description}
\item (a)
the sum of the potentials of S-comps of a C-comp, and of that C-comp,
is a multiple of $1/4$.
\item (b) a $s$-star or a $s$-comet has potential at least $-(5-s)/12$.
\end{description}

Invariant (a) follows from the fact that cost change is integral, and the other
potentials that change are edge potentials, each $1/4$.  Moreover, we coalesce
a group of C-comps if we charge more than one.  (A careful reasoning would
consider consequences of breaking a C-comp by applying a Path Step).

Invariant (b) is clearly true at the start of the process.  Then when we remove
an edge from an $s$ star we subtract 1 from $s$ and $1/12$ from the
potential, and the invariant is preserved.

\subsection{Preliminary to the Analysis of Phases 4-5}

When Phase 3 is over, we reorganize the potential in
a manner more appropriate for considering 3-stars.  We increase 
a potential of a C-comp from $-2/3$ to $-1/$, and we decrease
the potential of one of its S-comps.  In the same time, we want to
have the following potential for S-comps:

$$
p_4(S)=\left\{
\begin{array}{ll}
-\frac{1}{4} &
\mbox{if $S$ is a 3-star or a 3-comet} \\[2ex]
-\frac{3-s}{4} &
\mbox{if $S$ is an $s$-comet, $s<3$}
\end{array}
\right.
$$

Note that before the decrease, 3-stars
and 3-comets had potential at least $-2/12$, so it is OK.
Similarly, 1-comet had potential at least $-4/12$, so they
would get $-5/12>-2/4$, and the balance of 0 comets is even better.
We have a problem with 2-comets.  However, the reason is that we
want to make one increase in the entire C-comp.  If not even a single
S-comp has a ``slack'' to absorb this $1/12$ charge, then the sum of
the potentials in the C-comp is a multiple of $1/4$, plus $1/12$, and
this contradicts invariant (a).

Before phase 6 we need a different distribution of potential.  In that
phase we do not have 3-stars, the best cost index is above $1/2$.
The following values of potential are sufficiently high for the analysis:

$$
p_6(S)=\left\{
\begin{array}{ll}
-\frac{7}{12} &
\mbox{if $S$ is a 2-comet} \\[2ex]
-1 &
\mbox{if $S$ is a 1-comet} \\[2ex]
-\frac{29}{20} &
\mbox{if $S$ is a 0-comet} \\[2ex]
\end{array}
\right.
$$

Moreover, we will have potential zero for C-comps except for C-comps
that consist of one S-comp only; for such C-comp $C$ we can have
$p_6(C)=-1/3$ if $C$ is a 2-comet
and $p_6(C)=-1/4$ if $C$ is an 1-comet.

\subsection{Analysis of Phases 4-5}

In phase 4 we insert a maximum set of 3-stars to the solution, and this
selection can be modified in phase 5.  Note that we can obtain a set of
3-stars from the reference solution by taking all 3-stars and stripping
forks from 3-comets.

If a selected 3-star has terminals in three C-comps, we can collapse
it, $\Delta PE=\Delta PS=0$, $\Delta CA=3$, $\Delta CR=-4$ and
$\Delta PC=-1$, so this analysis step has balance zero.  And we still
can find at least as many 3-stars as we have 3-stars and 3-comets in
the reference solution.

Now consider connected component created by the inserted 3-stars together
with C-comps of the reference solutions, before we deleted any edges.
Suppose that this component had some $i$ C-comps, $j$ 3-stars and 3-comets
and within this component we inserted $j$ 3-stars (if we inserted more,
the balance is only more favorable).

In counting the balance, we have $\Delta CA=3i$, and we need to delete
$2i$ connections, so we split $\Delta CA$ into $2i$ equal parts to
analyze the balance of each connection deletion.  We can represent the
connections with artificial edges, and insert them one at the time, and
after each insertion remove a connection from the reference solution.

\noindent{\bf Case 1:} the new connection causes a deletion of
a non-edge.  This entails $\Delta CR=-2$ and $\Delta PC=1/2$, so we have
balance zero.  In the analysis of the subsequent cases we can
assume $\Delta PC=0$, because we consider Case 1 connections first,
with the exception, considered at the very end, that we can
annihilate each and every S-comp in the process, and thus increase $PC$
further by $1/2$.

\noindent{\bf Case 2:} the new connection causes a deletion of
a single connection from a 3-star.  This entails $\Delta CR=-1$,
$\Delta PE=-3/4$ and $\Delta PS=1/4$ for the balance of $-3/2+1+3/4-1/4=0$.

\noindent{\bf Case 3:} the new connection causes a deletion of
a single connection from a 3-comet.  This can be done in two ways: from
a fork or from the center to a terminal.  However, we can assume that no
3-star will survive the deletion process, otherwise they would be present
in the set we inserted.  Thus we alter a $(a,3)$-comet into $(a,2)$ comet.
We have $\Delta CR=-1$, $\Delta PE=-1/4$, $\Delta PS = -4/12$ for
the balance of $-3/2+1+1/4+4/12=1/12$.

\noindent{\bf Case 4:} the new connection causes a deletion of
a connection from a pre-existing 2-comet.

\noindent{\bf Case 4.1:} the deletion ``annihilates'' the 2-comet.  This
means that it was a (1,2)-comet.
We have $\Delta CR=-1$, $\Delta PE=-5/4$, $\Delta PS=1/4$ for the balance
of $-3/2+1+5/4-1/4=1/2$.

\noindent{\bf Case 4.2:} the deletion is from the center to a terminal,
so we change a $(a,2)$-comet to a $(a,1)$ comet.
We have $\Delta CR=-1$, $\Delta PE=-1/4$, $\Delta PS=-3/4$ for the balance
of $-3/2+1+1/4+3/4=1/2$.

\noindent{\bf Case 4.3:} the deletion is in a fork,
so we change a $(a,2)$-comet to a $(a-1,2)$ comet.
This entails $\Delta CR=-1$,
$\Delta PE=-3/4$ and $\Delta PS=-4/12$ for the balance
of $-3/2+1+3/4+4/12=7/12$.

\noindent{\bf Case 5:} the new connection causes a deletion of
a connection from a pre-existing 1-comet or 0-comet.  The balance
calculation is almost the same, except that
in the annihilation sub-case, $\Delta PE$ is more favorable, while in
other cases,
$\Delta PS$ is more favorable.

\noindent{\bf Case 6:} a subsequent deletion of a connection to
a deleted fork or an annihilated S-comp.  This entails $\Delta CR=-2$
with no changes in $PE$ and $PS$, for the balance of $-3/2+2=1/2$.
 
\noindent{\bf Case 7:} the new connection causes a subsequent deletion of
a connection from a pre-existing $b$-comet, $b\le 2$,
not annihilated by the first deletion.  In phase 6 we will have an
identical balance calculations, except for higher $\Delta CA$ because
of higher cost index of a collapsed comet.

\noindent{\bf Case 7.1:} the deletion ``annihilates'' a 2-comet.  This
means that it was a (1,2)-comet.
We have $\Delta CR=-1$, $\Delta PE=-5/4$, $\Delta PS=7/12$ for the balance
of $-3/2+1+5/4-7/12=1/6$.

\noindent{\bf Case 7.2:} the deletion ``annihilates'' a 1-comet.  This
means that it was a (2,1)-comet.
We have $\Delta CR=-1$, $\Delta PE=-7/4$, $\Delta PS=1$ for the balance
of $-3/2+1+7/4-1=1/4$.

\noindent{\bf Case 7.3:} the deletion ``annihilates'' a 0-comet.  This
means that it was a (3,0)-comet.
We have $\Delta CR=-1$, $\Delta PE=-9/4$, $\Delta PS=29/20$ for the balance
of $-3/2+1+9/4-29/20=3/10$.

\noindent{\bf Case 7.4:} the deletion ``reduces'' a 2-comet to a 1-comet.
This entails 
$\Delta CR=-1$, $\Delta PE=-1/4$, $\Delta PS=-5/12$ for the balance
of $-3/2+1+1/4+5/12=1/6$.

\noindent{\bf Case 7.5:} the deletion ``reduces'' a 1-comet to a 0-comet.
This entails 
$\Delta CR=-1$, $\Delta PE=-1/4$, $\Delta PS=-9/20$ for the balance
of $-3/2+1+1/4+9/20=4/20$.

\noindent{\bf Case 7.6:} the deletion removes a fork.
This entails 
$\Delta CR=-1$, $\Delta PE=-3/4$, $\Delta PS=0$ for the balance
of $-3/2+1+3/4=1/4$.

As we see, no deletion has negative balance.  However, we have
the problem of accounting for the remaining $p(C)=-1/2$.

This problem is solved is if the sum of balances
of the deletions was at least $1/2$. 

Thus we can inspect the possible combinations of deletions that do not
have such sum.  Clearly, they do not contain
any deletion in a pre-existing 2-, 1- or 0-comet,
nor a deletion in a fork or S-comp annihilated by
the first deletion (cases 4 and 5).

There are few cases of second deletions within 3-stars and 3-comets that
we need to consider.  First, a second deletion within an S-comp annihilated
by the first deletion is case 5, otherwise we have a combination of Case 3
with Case 7.1 or Case 7.4, with the balance of $1/4$.  Hence with two such
second deletions our problem is solved.

Thus if we had $i$ 3-stars and 3-comets, we made at most $i+1$ deletions
within them, hence $i-1$ other deletions, and of the latter, all were
Case 1, merging C-components.  Thus we started with as many C-comps
as we have 3-stars.  This also means that we had to have a second deletion
and the balance of the deletions is at least $1/4$.

Suppose that in the merged assembly of C-comps we had a $b$-comet, $b<3$;
we do not delete in that comet, so we can decrease its potential, and the
least decrease is $7/12-1/4=1/3$.  Together with the minimum balance of the
deletions, we have the desired $1/2$.

Now the picture becomes rather restricted.
In the reference solution
we have some $i$ C-comps, each consisting of exactly one S-comp, a 3-star or
a 3-comet.  We inserted a set of $i$ 3-stars, and each of them has at
least two terminals in one of C-comps, thus making one or two
connection inside and one or none outside.
Thus we have $i$ or $i-1$ connections made between C-comps.

If we have $i-1$ such connections, then they form a tree and we
can consider a leaf, one that does not contain an inserted 3-star.  
If this leaf is a 3-star, we can disregard it, because it means that
the inserted 3-star overlapping the leaf C-comp cannot create cycles.
If it is not a 3-star, in Phase 5 we can replace the inserted overlapping
3-star with a (1,3)-comet.

If we have $i$ such connections, then they form a simple cycle.  If there
is a comet on this cycle, we can make the same observation: replace one
of the overlapping 3-stars with a (1,3) comet without creating cycles.
If all C-comps are 3-stars, then we actually have balance.

Thus for every case when we have a deficit, the deficit is at most $1/4$
and we have an opportunity to replace a 3-star with a (1,3)-comet
in Phase 5.  Each such replacement improves the balance by at least 1,
hence it suffices to find $1/4$ fraction of the possible, and the greedy
selections obviously accomplish as much.

\subsection{Analysis of the Phase 6}

Basically, when we contract a comet with $i+1$ terminals, we have to
delete $i$ connections from the reference solution, and the accounting
is pretty much like in Case 7 in the previous sections, except that
we have a higher $\Delta CA$, which was in that calculation assumed to
be $3/2$.

If we remove from a 2-comet, then the portion of the cost per connection
is at most $5/3=3/2+1/6$, so it suffices that in the respective subcases
of Case 7 we had balance at least $1/6$.  Similarly, if we delete from a
1-comet, the cost per connection is at most $7/4=3/1/4$ and it suffices that
we had a balance of $1/4$ in those subcases.  For 0-stars, the cost
is at most $9/5=3/2+3/10$, and when we annihilate, we have such balance
in Case 7.3.  When we remove a fork without annihilation, then we have
at least (4,1)-comet and the cost is at most $12/7<3/2+1/4$, and the
balance of the fork removal is $1/4$.

We can formulate our conclusion as follows:
\begin{theorem}
The Six-Phase Algorithm has the approximation ratio of
$5/4$ for the Steiner Tree Problem in metrics with distances
1 and 2.
\end{theorem}

\hfill$\Box$

\end{document}